\documentclass[trackchanges,twocolumn,times]{aastex63}

\shorttitle{Loop magnetic field measurements}                 

\shortauthors{Brooks et al. }
\journalinfo{To be submitted to the Astrophysical Journal}

\usepackage{grffile}

\begin{document}

%% ------------------------------------------------------------------------------------------
%% --- TITLE PAGE ---------------------------------------------------------------------------
%% ------------------------------------------------------------------------------------------

\title{Measurements of Coronal Magnetic Field Strengths in Solar Active Region Loops}

\author[0000-0002-2189-9313]{David H.\ Brooks}
\affil{College of Science, George Mason University, 4400 University Drive, Fairfax, VA 22030, USA}
\affil{Hinode Team, ISAS/JAXA, 3-1-1 Yoshinodai, Chuo-ku, Sagamihara, Kanagawa 252-5210, Japan}

\author[0000-0001-6102-6851]{Harry P. Warren}
\affil{Space Science Division, Naval Research Laboratory, Washington, DC 20375, USA}

\author[0000-0002-9325-9884]{Enrico Landi}
\affil{Department of Climate and Space Sciences and Engineering, University of Michigan, Ann Arbor, MI, USA}

%% ------------------------------------------------------------------------------------------
%% --- ABSTRACT -----------------------------------------------------------------------------
%% ------------------------------------------------------------------------------------------

\begin{abstract}
The characteristic electron densities, temperatures, and thermal distributions of 1\,MK\, active region loops 
are now fairly well established, but their coronal magnetic field strengths remain undetermined. 
Here we present measurements from a sample of coronal loops observed by the Extreme-ultraviolet Imaging
Spectrometer (EIS) on {\it Hinode}. We use a recently developed diagnostic technique that involves atomic
radiation modeling of the contribution of a magnetically induced transition (MIT) to the \ion{Fe}{10} 257.262\,\AA\, 
spectral line intensity. We find coronal magnetic field strengths in the range of 60--150\,G. We discuss some aspects
of these new results in the context of previous measurements using different spectropolarimetric techniques, and
their influence on the derived Alfv\'{e}n speeds and plasma $\beta$ in coronal loops.
\end{abstract}

\keywords{Solar coronal loops: Magnetic fields: Solar extreme ultraviolet emission}

%% ------------------------------------------------------------------------------------------
%% --- BODY ---------------------------------------------------------------------------------
%% ------------------------------------------------------------------------------------------

\section{Introduction}

The magnetic field in the Sun's corona is the primary driver of activity. To fully understand the processes
that heat the solar corona, drive flares and coronal mass ejections (CME), and form and accelerate the solar wind, we need to fully establish the coronal
magnetic structure. Underpinning much of coronal physics research are models that extrapolate the photospheric
magnetic field. We expect significant advances in our understanding of all aspects of coronal structure and dynamics
would be made if we could reliably measure the actual topology and evolution of the coronal magnetic field.

Indirect measurements of coronal magnetic field strengths have been attempted through novel techniques such as 
coronal seismology \citep{Nakariakov2001,Vandoorsselaere2008}, or determining the geometric properties of a CME
flux rope and leading shock \citep{Gopalswamy2012}. Attempts have also been made to infer the coronal magnetic field
using nonlinear force-free field (NLFFF) extrapolations \citep{DeRosa2009,Malanushenko2009,Malanushenko2012}.
These efforts highlight that direct measurements are difficult to make.
Attempts have been made using Stokes V circular polarization measurements in a variety of spectral lines \citep{Lin2000,Kuridze2019},
microwave spectropolarimetry \citep{Fleishman2020}, and techniques that combine plasma density measurements with the phase
speed of magnetohydrodynamic (MHD) waves \citep{Yang2020}. These studies have generally found coronal magnetic field 
strengths of a few G\, below 1.4 solar radii, and up to 50\,G\, in active region loops. Higher field strengths of a few hundred
\,G\, have been measured in flares and post-flare loops.
The Daniel K. Inouye Solar Telescope \citep[DKIST,][]{Rimmele2020} is now providing
a new capability for measuring coronal magnetic fields.

Recently, a new direct measurement technique has been developed that does not depend on spectropolarimetry. The method relies
on a magnetically induced transition (MIT) that contributes
to the intensity of the Fe X 257.262\,\AA\, spectral line in the presence of an external magnetic field. With no
magnetic field, the line is formed as a combination of emission from an electric dipole (E1) transition from the 3s$^2$3p$^4$3d $^4$D$_{5/2}$
level, and a magnetic quadrupole (M2) transition from the 3s$^2$3p$^4$3d $^4$D$_{7/2}$ level, both down to the
3s$^2$3p$^5$ $^2$P$^o_{3/2}$ ground state. The Zeeman effect
due to the magnetic field mixes the $^4$D$_{5/2}$ and $^4$D$_{7/2}$ levels, and opens
a new pathway to the ground state. The idea was first proposed by \citet{Li2015} and discussed
further by \citet{Li2016} and \citet{Judge2016}. 
\citet{Si2020} and \citet{Landi2020} subsequently developed techniques to enable measurements using observations by the 
{\it Hinode} \citep{Kosugi2007} EUV Imaging Spectrometer \citep[EIS,][]{Culhane2007}.
The methods have been applied to EIS observations of the active region corona
\citep{Si2020,Landi2020}, a C-class flare \citep{Landi2021}, and the footpoints of active region core loops \citep{Brooks2021}, but these
developments now allow use of the large untapped resource that is the EIS post-launch database (14.5 years at the time of writing).

Active region loops formed near 1--2\,MK are of great interest because of their intriguing properties. These loops
persist longer than expected cooling times, are overdense compared to static equilibrium theory \citep{Aschwanden2000,Winebarger2003}, 
have narrow, but not quite isothermal, temperature
distributions \citep{Warren2008}, and are close to being spatially resolved \citep{Brooks2012}. An extensive review of their properties
can be found in \citet{Reale2014}. The         
coronal magnetic field strengths in these loops, however, are currently not known accurately.

When measuring their properties,
it is critical to account for line-of-sight contributions from background/foreground emission \citep{DelZanna2003}. This line-of-sight
superposition is even more difficult to deal with in off-limb measurements, such as would be made with ground telescopes, because of the
longer path lengths. The EIS capability
to measure magnetic field strengths in coronal loops observed on disk allows us to deal meaningfully with this problem.

Here we measure the coronal magnetic field strength in a sample of active region loops formed near 1\,MK. We extract
co-spatial background subtracted intensities for 18 loop segments that are prominent in Fe X and Fe XI
emission lines. We then determine their magnetic field strengths.
Note that the loops that are bright in Fe X and Fe XI are usually high lying structures towards the active
region boundary, so there is some overlap with what the community would usually call `fan' loops as well as `warm' loops.

\section{Observations and data analysis}
\label{data}
{\it Hinode}/EIS records EUV spectra in two wavelength
intervals covering 171--211\,\AA\, and 245--291\,\AA\, with a spectral resolution of 23\,m\AA\, and there are four slit options
(1$''$, 2$''$, 40$''$, and 266$''$). The data we use were reduced 
and processed using the standard procedure eis\_prep from the SolarSoftware IDL library \citep{Freeland1998} -- current as of April 2021. The calibration software treats
instrumental issues such as the effects of contaminated pixels and cosmic ray strikes. It also converts the raw data to
physical units (erg cm$^{-2}$ s$^{-1}$ steradian$^{-1}$) for analysis. We use the artificial neural network model of
\citet{Kamio2010} to correct positional offsets between the two EIS CCDs, and the spectral drift due to orbital thermal
effects.

We measure the coronal magnetic field strength using a technique that extracts the excess emission in the Fe X 257.262\,\AA\, spectral line
due to the production of the MIT by the coronal magnetic field. A method to utilise this line was first investigated by
\citet{Si2020}. Their technique relied on using the Fe X 174.532\,\AA\, spectral line ratioed with Fe X 257.262\,\AA, although unfortunately this ratio
is only weakly sensitive to the magnetic field strength (about 15\% over 1--1000\,G). The Fe X 174.532\,\AA\, line is also rather
weak itself. \citet{Landi2020} therefore further
developed this technique, by estimating the contributions of the E1 and M2 transitions to the Fe X 257.262\,\AA\, line,
using the much stronger Fe X 184.536\,\AA\, spectral line.
When the magnetic field strength does not exceed $\sim$200\,G, the MIT intensity can be retrieved by removing the E1 and M2 contributions 
determined by the intensity ratios with the Fe X 184.536\,\AA\, line under the assumption of zero magnetic field; 
the branching ratio between the MIT and M2 transitions can therefore be determined. This ratio
has a much stronger dependence on the magnetic field strength, and can be determined from
\begin{equation}
{I_{MIT} \over I_{M2}} = { I_{257} - (r(E1M2/184) \times I_{184}) \over r(M2/184) \times I_{184}}
\label{equation}
\end{equation}
where $I_{257}$ and $I_{184}$ are the measured intensities of Fe X 257.262\,\AA\, and Fe X 184.536\,\AA, respectively,
$r(E1M2/184)$ is the ratio of the theoretical magnetic field-free E1 + M2 contribution to Fe X 257.262\,\AA\, to the Fe X 184.536\,\AA\, intensity,
$r(M2/184)$ is the ratio of the theoretical magnetic field-free M2 contribution to Fe X 257.262\,\AA\, to the Fe X 184.536\,\AA\, intensity,
$I_{MIT}$ is the inferred MIT contribution to the intensity of Fe X 257.262\,\AA\, (upper line in the equation), and 
$I_{M2}$ is the inferred intensity of the M2 transition (lower line in the equation). The theoretical calculations were
made using the CHIANTI v.10 database \citep{DelZanna2021}.

This method corresponds to the weak magnetic field technique discussed by \citet{Landi2020}, and is valid for magnetic field strengths 
below $\sim$200\,G. For higher field strengths, the population of the $^4$D$_{7/2}$ level becomes sensitive to the magnetic field, 
so that the M2 intensity can not be directly estimated through zero-field ratios with the Fe X 184.536\,\AA\, line, and another method
involving the ratio of the total M2 + MIT intensity to the Fe X 184.536\,\AA\, intensity is preferred. For all the fine details of these 
techniques we refer to the papers cited above and in the introduction.

Note that to compute the theoretical ratios we also have to measure the electron density. We used the Fe XI 182.167/(188.299+188.216)
density sensitive ratio suggested by \citet{Landi2020}. One consequence is that we have to ensure the loops we select are the same when observed
in Fe X or XI. 
This is a condition that limits our selection. The Fe XI lines, however, are stronger
than the Fe X density sensitive ratios that involve the weak Fe X 174.532\,\AA\, line.

For this analysis, we focus on large field-of-view (FOV) scans of two active regions taken in 2007, May and December. 
EIS generally downloads only a subset of all possible spectral lines in each observation. This is because of limitations
on the downlink telemetry. Prior to early 2008, however, the X-band transmission system was fully functional, so larger
volumes of data could be generated. This means that 2007 is a rich period for data mining of observations with extended wavelength 
coverage. This is helpful for finding EIS observations that contain the critical Fe X 257.262\,\AA\, line. Furthermore, perhaps the biggest
uncertainty in applying the EIS magnetic field strength measuring technique is the fact that it relies on spectral lines that
fall on different detectors, and therefore depends on the detector cross-calibration. This is known to have evolved over time,
but there is not yet a consensus on the best correction method \citep[see][]{DelZanna2013,Warren2014}. By using observations
taken early in the mission, we avoid the need to consider the time-dependence of the calibration.

\begin{figure*}
\centering
\includegraphics[width=1.0\textwidth]{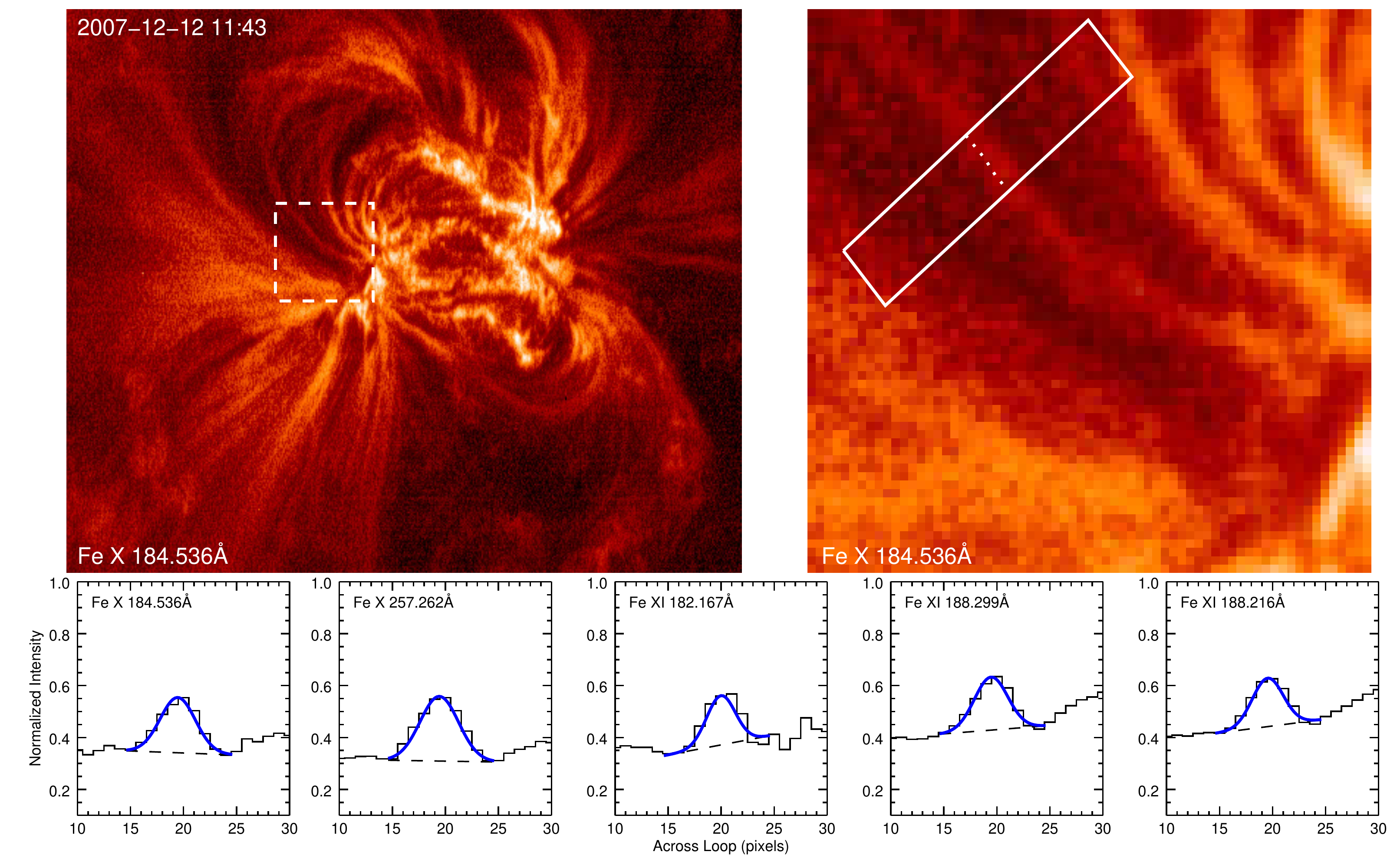}
\caption{ Example showing one of the selected loops. Top left: EIS Fe X 184.536\AA\, intensity image of AR 10978 on 2007, December 12. 
The white dashed box shows the region
corresponding to the image on the top right. Top right: zoomed image of the area surrounding the loop. The white dotted line and box show the loop segment chosen
to prepare the cross-field intensity profile. Bottom row: cross-field intensity profiles for the Fe X magnetic field and Fe XI density diagnostic spectral lines.
The histogram shows the data normalized by the peak intensity across the profile. The background is shown by the dashed line, and a Gaussian fit to the 
profile is shown by the blue line. The interpolated data (see text) have been resampled to the instrument pixel scale. The images in the top row were
treated with the multi-scale Gaussian normalization algorithm of \citet{Morgan2014}.
}
\label{fig1}
\end{figure*}

\begin{figure}[h]
\centering
\includegraphics[width=0.5\textwidth]{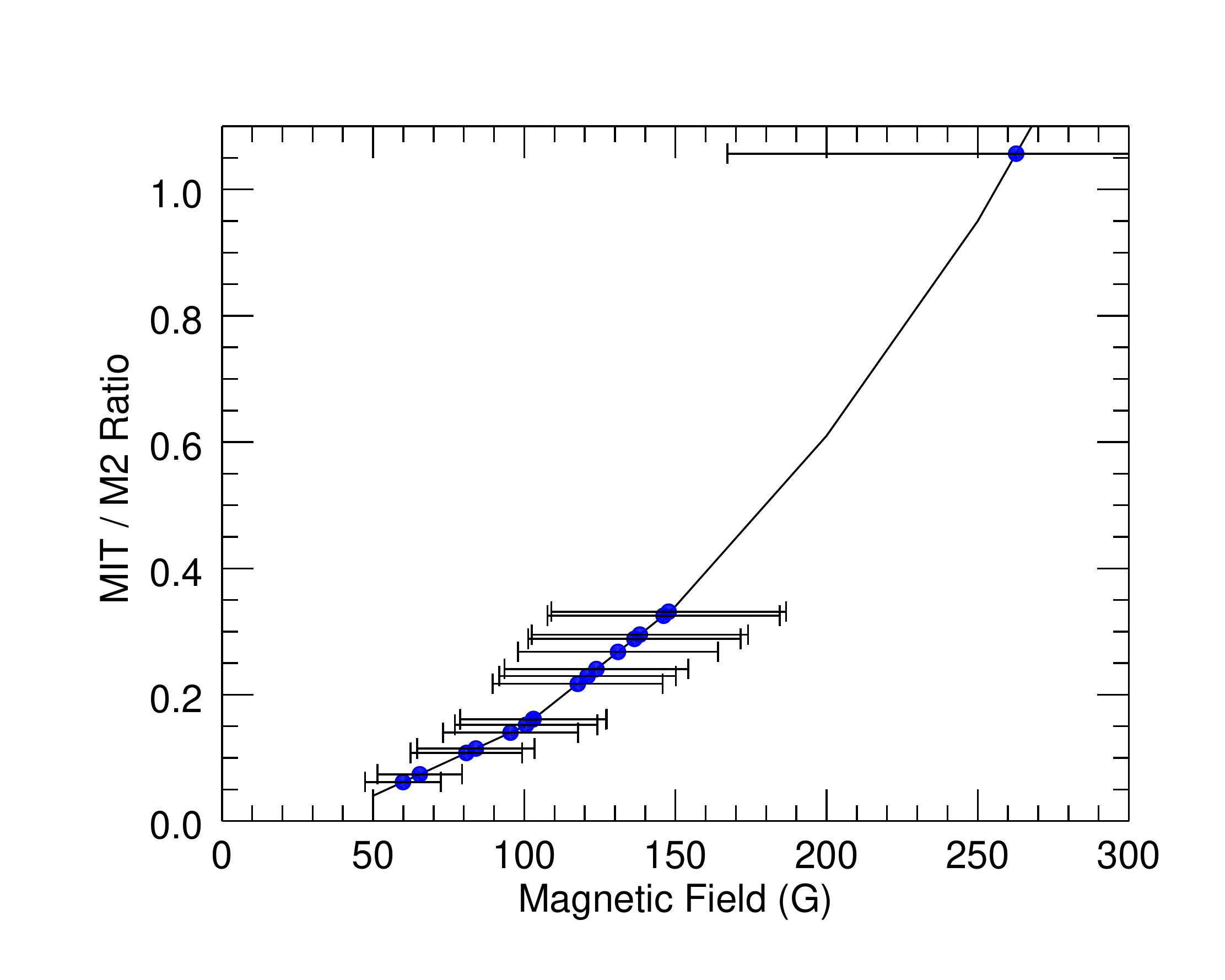}
\caption{ Measurements of the loop segment magnetic field strength (blue dots) overlaid on the theoretical curve of the MIT/M2 intensity ratio as a function of field strength.
}
\label{fig2}
\end{figure}

\begin{figure}[h]
\centering
\includegraphics[width=0.5\textwidth]{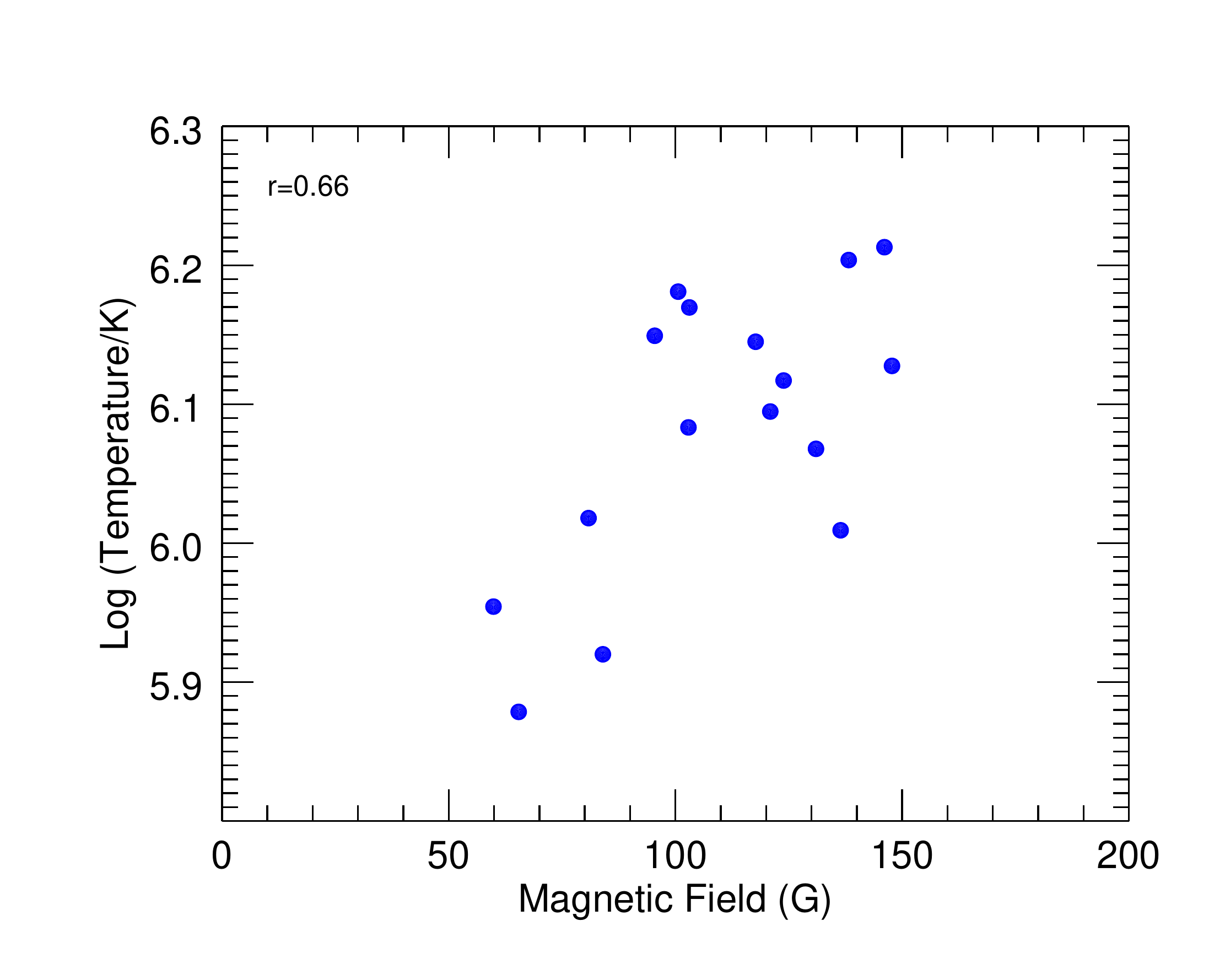}
\caption{ Measurements of the magnetic field strength plotted against the logarithm of the temperature for each loop segment. $r$ is the linear Pearson
correlation coefficient.
}
\label{fig3}
\end{figure}

\begin{figure}[h]
\centering
\includegraphics[width=0.5\textwidth]{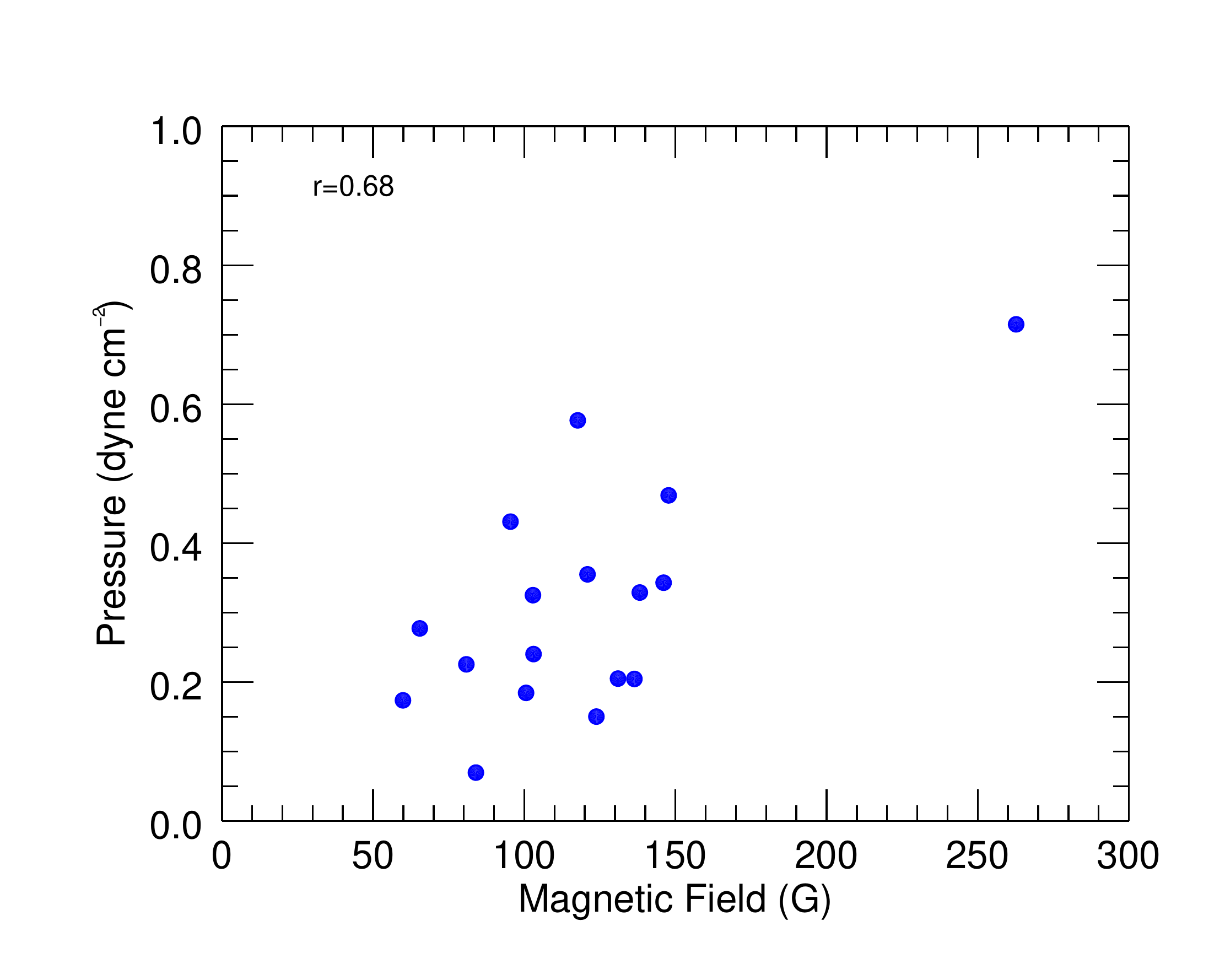}
\caption{ Measurements of the magnetic field strength plotted against the dynamic pressure for each loop segment. $r$ is the linear Pearson
correlation coefficient.
}
\label{fig4}
\end{figure}

\begin{deluxetable*}{lccccccccccccc}
\tabletypesize{\small}
\tablecaption{Coronal magnetic field measurements of active region loops observed by EIS}
\tablehead{
\multicolumn{1}{c}{No.} &
\multicolumn{1}{c}{Date} &
\multicolumn{1}{c}{Time} &
\multicolumn{1}{c}{$n$} &
\multicolumn{1}{c}{$T$} &
\multicolumn{1}{c}{I$_{184}$} &
\multicolumn{1}{c}{I$_{257}$} &
\multicolumn{1}{c}{E1M2/184} &
\multicolumn{1}{c}{M2/184} &
\multicolumn{1}{c}{MIT/M2} &
\multicolumn{1}{c}{B} &
\multicolumn{1}{c}{v$_A$} &
\multicolumn{1}{c}{c$_s$} &
\multicolumn{1}{c}{$\beta$} 
}
\startdata
01 & 05/01 & 02:57 & 9.3 & 6.1 & 750$\pm$160 & 530$\pm$120 & 0.64 & 0.35 & 0.21$\pm$0.11 & 120$\pm$28 & 5800 & 200 & 1.2(-3)\\
02 & 05/01 & 02:01 & 9.2 & 6.0 & 670$\pm$150 & 490$\pm$110 & 0.70 & 0.41 & 0.10$\pm$0.05 & 81$\pm$18 & 4500 & 190 & 1.7(-3)\\
03 & 05/19 & 13:16 & 9.2 & 6.0 & 5500$\pm$1200 & 4600$\pm$1000 & 0.71 & 0.43 & 0.27$\pm$0.14 & 140$\pm$36 & 7800 & 180 & 5.5(-4)\\
04 & 12/10 & 03:20 & 9.3 & 6.1 & 590$\pm$130 & 420$\pm$93 & 0.66 & 0.36 & 0.15$\pm$0.08 & 100$\pm$23 & 5100 & 200 & 1.5(-3)\\
05 & 12/10 & 03:28 & 9.4 & 5.9 & 2200$\pm$490 & 1400$\pm$300 & 0.60 & 0.29 & 0.07$\pm$0.03 & 65$\pm$13 & 2800 & 160 & 3.2(-3)\\
06 & 12/12 & 15:08 & 8.9 & 6.1 & 320$\pm$70 & 300$\pm$66 & 0.82 & 0.55 & 0.22$\pm$0.11 & 120$\pm$29 & 9400 & 210 & 4.9(-4)\\
07 & 12/12 & 14:48 & 9.1 & 6.0 & 510$\pm$110 & 380$\pm$84 & 0.72 & 0.43 & 0.06$\pm$0.03 & 60$\pm$12 & 3500 & 170 & 2.4(-3)\\
08 & 12/12 & 14:47 & 9.4 & 6.1 & 650$\pm$140 & 460$\pm$100 & 0.60 & 0.30 & 0.31$\pm$0.16 & 150$\pm$39 & 6400 & 210 & 1.0(-3)\\
09 & 12/12 & 14:40 & 9.2 & 6.2 & 310$\pm$68 & 260$\pm$56 & 0.71 & 0.42 & 0.28$\pm$0.14 & 140$\pm$36 & 7800 & 230 & 8.6(-4)\\
10 & 12/12 & 12:58 & 8.8 & 5.9 & 1000$\pm$220 & 960$\pm$210 & 0.88 & 0.62 & 0.11$\pm$0.05 & 84$\pm$19 & 7400 & 170 & 4.9(-4)\\
11 & 12/13 & 16:08 & 9.1 & 6.1 & 440$\pm$97 & 380$\pm$83 & 0.74 & 0.46 & 0.25$\pm$0.13 & 130$\pm$32 & 8000 & 200 & 6.0(-4)\\
12 & 12/13 & 15:36 & 9.1 & 6.2 & 790$\pm$170 & 650$\pm$140 & 0.76 & 0.47 & 0.15$\pm$0.08 & 100$\pm$23 & 6500 & 220 & 1.1(-3)\\
13 & 12/13 & 15:53 & 9.8 & 5.9 & 1000$\pm$230 & 590$\pm$130 & 0.45 & 0.13 & 0.99$\pm$0.50 & 260$\pm$94 & 6800 & 160 & 5.2(-4)\\
14 & 12/15 & 20:06 & 9.2 & 6.2 & 870$\pm$190 & 720$\pm$160 & 0.70 & 0.41 & 0.31$\pm$0.16 & 150$\pm$39 & 8200 & 230 & 8.0(-4)\\
15 & 12/15 & 20:12 & 9.5 & 6.1 & 720$\pm$160 & 460$\pm$100 & 0.58 & 0.27 & 0.20$\pm$0.10 & 120$\pm$28 & 4700 & 220 & 2.0(-3)\\
16 & 12/15 & 20:08 & 9.3 & 6.1 & 1100$\pm$250 & 760$\pm$170 & 0.63 & 0.33 & 0.13$\pm$0.07 & 95$\pm$22 & 4400 & 220 & 2.3(-3)\\
17 & 12/18 & 02:07 & 8.9 & 6.2 & 700$\pm$160 & 630$\pm$140 & 0.81 & 0.54 & 0.15$\pm$0.07 & 100$\pm$23 & 7400 & 230 & 9.1(-4)  
\enddata
\tablenotetext{}{All dates are in 2007. Electron densities ($n$) and temperatures ($T$) are base-10 logarithms in units of cm$^{-3}$ and $K$. 
I$_{184}$ and I$_{257}$ are the Fe X 184.536\,\AA\, and Fe X 257.262\,\AA\, intensities, respectively, and are in units of erg cm$^{-2}$ s$^{-1}$ steradian$^{-1}$.
E1M2/184 is the modeled contribution of the E1 and M2 transitions to I$_{257}$ normalized by I$_{184}$. 
M2/184 is the modeled contribution of the M2 transition to I$_{257}$ normalized by I$_{184}$. 
MIT/M2 is the MIT to M2 intensity ratio calculated from Equation \ref{equation}. 
Coronal magnetic field strengths ($B$) are in units of Gauss. Alfven ($v_A$)
 and sound ($c_s$) speeds are in units of km s$^{-1}$. The exponents of the plasma beta ($\beta$) are shown in brackets.
 Values are reported to two significant figures. }
\label{table}
\end{deluxetable*}

Figure \ref{fig1} shows an example Fe X 184.536\,\AA\, raster image of AR 10978 taken on 2007, December 12. For
this observation, the 1$''$ slit was used to scan an FOV of 460$''\times$384$''$ and the exposure time at each slit position was 40\,s. We were able to collect a sample
of 15 loops from observations of this AR as it crossed the disk from December 10 to 18 -- but see the discussion below. To supplement the sample,
we added 3 more loops from AR 10953 and 10956 that were observed on May 1 and 19. The raster used for those observations was an early
version of the same one, with the same exposure time but a slightly smaller FOV of 330$''\times$304$''$.

A critical component of reliable loop intensity measurements is an adequate background removal technique. Background/foreground emission
is known to contribute significantly along the line-of-sight \citep{Klimchuk1992,DelZanna2003,Aschwanden2005,Reale2014,Brooks2019}, and can distort
the measured properties. Here we use a well-established technique introduced by \citet{Aschwanden2008}, that we have developed and made use of
in several previous studies \citep{Warren2008,Brooks2012,Brooks2013}. We select a loop segment that is clearly visible in Fe X 184.536\,\AA\, 
spectral images. An example is shown in the upper right panel of Figure \ref{fig1}. The Fe X 184.536\,\AA\, spectral image is formed by fitting a 
single Gaussian to the intensity profile at each pixel. We used a single Gaussian for all the spectral lines analyzed in this study. We extract cross-loop
intensity profiles perpendicular to the loop axis and at several positions within the selected segment (boxed region). These are resampled to a higher resolution (sub-pixel) scale.
We then average these profiles along the loop axis.

Examples of the cross-loop averaged intensity profiles are shown in the lower panel of Figure \ref{fig1} for all the lines needed for the magnetic
field measurements. The histograms show the data (resampled back to the EIS pixel scale). We subtract the loop co-spatial background emission by 
visually selecting two background positions and fitting a first order polynomial between them. This is the dashed line in the plots of Figure \ref{fig1}.
The background positions were selected using the Fe X 184.536\,\AA\, profile and the same pixels were used for all the lines. After background subtraction,
we fit a Gaussian function to the remaining profile and the area of the Gaussian is the loop intensity.

We also obtain a measure of the electron temperature for each loop. 
Since 1\,MK\, active region loops typically have narrow temperature distributions \citep{Warren2008}, it suffices to assume a Gaussian emission measure
function and find the fit that best reproduces the
background subtracted intensities to localize the peak (we are not concerned here with the shape of the distribution). We then take the temperature of
the peak of the emission measure distribution as the temperature of the loop segment. To do this, we add background-subtracted intensities
from spectral lines from Mg V--VII, Si VII, Si X, and Fe VIII--XVI ions, which cover a wide, and adequate, range of temperatures ($\log (T/K)$ = 0.3--2.8\,MK). 
The co-spatial background was subtracted using the same method as for Fe X.
Recall that we ensured that each loop observed in Fe X was also observed in Fe XI to use the density diagnostic ratio. To do this we only included examples
where the cross-field loop intensity profiles were highly correlated ($r>$0.8) with the Fe X cross-field intensity profile. 
We apply the same criteria here for the emission measure analysis, and include
a 22\% uncertainty to account for the intensity calibration error \citep{Lang2006}. If the cross-field intensity profile for a particular spectral line
was not correlated with the Fe X cross-field intensity profile we assume that the emission does not come from the same loop and 
set the observed intensity to zero. In these cases we take the error as 20\% of the background intensity. We also 
use the electron density we previously calculated to compute the theoretical contribution functions needed for the emission measure 
measurements. 

Having obtained measurements of the loop temperature, electron density, and coronal magnetic field strength, we are also able to compute several
plasma quantities for comparisons with theoretical models. We therefore calculated the coronal Alfv\'{e}n speed, $v_A$, sound speed, $c_s$, and ratio
of plasma pressure to magnetic pressure, $\beta$, as follows
\begin{equation}
v_A = {B \over \sqrt{4 \pi \mu m_p n}} \hspace{0.4cm} ; \hspace{0.4cm} c_s = \sqrt{ {\gamma P \over m_p n}} \hspace{0.4cm} ; \hspace{0.4cm} \beta = { 8 \pi P \over B^2 }
\label{plasma_equation}
\end{equation}
where $\mu$ is the permeability of free space, $m_p$ is the proton mass, $n$ is the electron density, $\gamma$ is the ratio of specific heats, $P$ is the plasma pressure, and $B$ is the magnetic field strength.

\section{Results and discussion}
We summarize the results of our analysis in Table \ref{table} and Figure \ref{fig2}. The table gives the date of each
loop measurement, and the time when the EIS slit was scanning over the position of the selected loop segment. The electron
density was calculated using the Fe XI ratio, and the temperature was calculated from the Gaussian emission measure analysis. We provide the rest of the parameters needed to use Equation \ref{equation}
for transparent reproduction of our results.
The loop intensities in Fe X 184.536\,\AA\, and Fe X 257.262\,\AA. The theoretical E1 and M2 transition contributions to 
Fe X 257.262\,\AA\, divided by the Fe X 184.536\,\AA\, intensity, $r(E1M2/184)$. The theoretical M2 only contribution to 
Fe X 257.262\,\AA\, divided by the Fe X 184.536\,\AA\, intensity, $r(M2/184)$. The MIT/M2 intensity ratio, $I_{MIT}/I_{M2}$ 
computed using Equation \ref{equation}. And finally the magnetic field strength, $B$. The latter is derived from the 
MIT/M2 ratio as a function of magnetic field strength (Figure \ref{fig2}). 
The table also shows our computed values of the Alfv\'{e}n and sound speed, and the plasma $\beta$.

The mean electron density for the loops in our sample is $\log$ (n$_e$/cm$^{-3}$) = 9.2, which is typical for `warm' 1--2\,MK\, active region loops. 
The actual mean temperature of the loops is 1.1\,MK.
The mean magnetic field strength is $\sim$110\,G, with the sample of loops falling in the range of 60--150\,G. These
values are plotted as blue dots in Figure \ref{fig2}. 
In one exceptional case the field strength is rather high: 260\,G.
As discussed in Section \ref{data}, we used the weak field approximation for this study. This is valid for field strengths up to $\sim$200\,G\, whereupon the strong field approximation should be used \citep{Landi2020}. Adopting the strong field approximation, the value is within a few percent for this loop. This is an encouraging sign that the two methods work consistently.

Alfv\'{e}n speeds fall in the range of $\sim$2000--9400\,km s$^{-1}$; much larger than the local coronal sound speeds ($\sim$160--230\,km s$^{-1}$).
The mean plasma beta for this loop sample is $\sim$0.002. 

We did not find any obvious relationship between our measured coronal loop magnetic field strengths and the magnitude of the photospheric magnetic flux
of the active regions in this survey.
We performed some preliminary checks, however, to investigate whether we could uncover any obvious correlations with
other properties of the loops. Excluding the loop segment where the measured field strength was too high
for the weak field approximation to be valid, we found a strong correlation ($r$=0.66) with the measured temperature, 
such that higher magnetic field strengths lead to higher loop temperatures. We show this relationship in Figure \ref{fig3}. 
This could be influenced by the location where the measurements were made around the loop, although our sample does cover quite a
wide range of loop properties from cooler, denser cases to hotter, more tenuous ones. We also found that a relationship can be 
established with the loop pressure that extends to the loop segment with the highest field strength (shown in Figure \ref{fig4}). This
segment also has the highest density and lowest temperature, which act to offset each other and allow the loop to follow the same trend as the others
and the correlation between the parameters is also strong ($r$=0.68). Our initial survey has been limited by the need to 
select a sample from data taken early in the mission when calibration degradation was minimal. Ongoing progress in improving the
EIS calibration will allow a wider study to take advantage of more extensive data (including flare data) in the future. This should help to clarify these
relationships further.

\cite{Landi2020} estimate the uncertainty in their measurements to be on the order of 70\% -- driven by multiple factors
such as the cross-calibration between the short- and long-wavelength detectors, the density diagnostic ratio,
atomic physics parameters, and also the accuracy of the energy separation between the $^4$D$_{5/2}$ and $^4$D$_{7/2}$ levels.
This includes an estimate of the contribution from the radiometric calibration and sensitivity degradation over time.
As discussed, we chose observations from early in the mission to minimize the impact of the time-dependence of the calibration.
Based on the \citet{Warren2014} model, sensitivity loss is minimal ($<$10\%) for the datasets we use, so this contribution is 
reduced from 50\% to 24\%. This propagates through to produce an uncertainty in $I_{MIT}/I_{M2}$ of 51\%. 
We adopt this value in Table \ref{table}. Furthermore, the conversion from $I_{MIT}/I_{M2}$ to $B$ is not linear (see Figure \ref{fig2}).
To assess the impact of the uncertainties on 
this conversion, for each loop, we performed a 10000 run Monte-Carlo simulation where we randomly perturbed the $I_{MIT}/I_{M2}$ value within a uniform distribution covering the range defined by the 51\% uncertainty.
We then take the standard deviation of the magnetic field strength distribution, with the estimated $\sim$20\% 
uncertainty in the $^4$D$_{5/2}$ and $^4$D$_{7/2}$ energy separation added in quadrature, as our evaluation of the uncertainty in $B$. These range
from 20--36\% and are added in Table \ref{table}. 

These uncertainties raise an important issue. The measurements are difficult and the methodology we use requires good data in several aspects.
For example, while background subtraction has proven to be neccesary for accurately determining loop properties, if the loop is 
embedded close to other distinct loops -- such as in the fans -- then the overlap of emission along the line-of-sight could be overestimated. 
This is clearly not the case for the isolated loop in Figure \ref{fig1}, but it highlights that the loops need to be reasonably prominent above the 
background emission to make the measurement.
When the measured field was negative we discarded the loop under investigation. This can happen if the modeled E1+M2 transition contribution
to Fe X 257.262\,\AA\, is relatively high, or if the $I_{257}/I_{184}$ ratio is relatively low (see equation \ref{equation}). In these cases
the $I_{MIT}/I_{M2}$ ratio becomes negative and 
the magnetic field strength is too low to be measured even by the weak field technique.
We also discarded any loops with field strengths lower than the measurable limit
of the method \citep[$\sim$50\,G,][]{Landi2020}
To be clear, this is a general problem in spectroscopic analysis and not necessarily specific to the magnetic
field measurements: we also discarded loops where the measured density was negative. It may lead, however, to a selection effect towards 
loops with strong enough field strength in our sample. The lowest field strength we were able to measure was 59\,G.
To gain some insight into this problem, we recorded the reason we rejected each loop from our analysis.
In the end, we discarded a comparable number of loop segments from further analysis for this reason. This is why we finally added more
loops from ARs 10953 and 10956.

Our magnetic field strength measurements are higher than reported previously in off-limb observations of the corona obtained from ground telescopes. 
\citet{Yang2020}, for example, measured values of 1--4\,G\, at heights ranging from 1.05--1.35 solar radii. We mentioned earlier
that the path integration length along the line-of-sight at the limb results in a larger contribution from the background/foreground
emission. It could be that our field strengths are higher because we remove the co-spatial background around the loop. To test
this explanation, we re-computed the field strengths for the loops in our sample retaining the background emission. In this case our field strengths 
are indeed lower, with a median value of 19\,G. The reason is that while the background emission may have comparable levels of magnetically insensitive E1 and M2 intensity, 
it likely has a lower magnetic field strength. This leads to the magnetic signature of the loop being diluted, and appearing to be a structure with a lower magnetic field. This result also gives an indication of the effect of overestimating the background. Because of the limitations in the sensitivity of the technique just discussed,
however, many of the field strength measurements become negative. It would be better to gather a dedicated sample focused on this
particular measurement before drawing a definitive conclusion.

We mentioned selection effects: to be clear, this is a general problem in spectroscopic analysis and not necessarily specific to our magnetic
field measurements. Coronal seismology techniques, for example, require loop waves/oscillations to make the measurement. This also suggests that there
could be real physical reasons for our higher field strengths. Perhaps a lower confining field strength, and potentially less field line braiding,
leads to a higher prevalence of loop oscillations.

Given that our magnetic field strengths are higher, and considering that $v_A$ depends linearly on $B$, it is not surprising
that our calculated Alfv\'{e}n speeds are fairly high. Similarly, since $\beta$ depends on the inverse square of $B$, it is also
not surprising that the plasma beta for the loops in this sample is small. Our values correspond to approximately the low- to middle-corona
in the plasma $\beta$ model of \citet{Gary2001} i.e. below $\sim$100\,Mm. This might anyway be reasonable since the measurements are made using
Fe X. 

It will be interesting to compare actual EIS loop field strength measurements with magnetic extrapolations from the photosphere.
\citet{Landi2020} compared EIS off-limb measurements with the radial decrease of magnetic field strength obtained from a potential field source surface extrapolation.
They found that the measured field falls more slowly -- due to a combination of the likely presence of non-potential fields and the weakness of 
the Fe X spectral lines high off-limb. Comparisons with individual loops are more challenging.
Typically the observed intensities scale to some power of $\bar{B}/L$, where $\bar{B}$ is the average magnetic field strength along the loop
and $L$ is the loop
length. \citet{Ugarte2019}, for example, found that Fe XVIII intensities in active region loops scale as $(\bar{B}/L)^{1/2}$. For the longer,
cooler loops in our sample we would expect lower heating rates unless the field is very strong. The mean field strength in the extrapolations
in \citet{Ugarte2019}, however, is about a factor of two higher than we measure here. 

%% ------------------------------------------------------------------------------------------
%% --- ACKNOWLEDGMENTS ----------------------------------------------------------------------
%% ------------------------------------------------------------------------------------------

\acknowledgments 
The work of DHB and HPW was funded by the NASA Hinode program. The work of EL is funded by NASA grants 80NSSC18K1208, 80NSSC20K0185 and NSF grant AGS-1408789. Hinode is a Japanese mission developed and launched by ISAS/JAXA, with NAOJ as domestic partner and NASA and STFC (UK) as international partners. It is operated by these agencies in co-operation with ESA and NSC (Norway). 
CHIANTI is a collaborative project involving George Mason University, the University of Michigan (USA), University of Cambridge (UK) and NASA Goddard Space Flight Center (USA).

%% ------------------------------------------------------------------------------------------
%% --- REFERENCES ---------------------------------------------------------------------------
%% ------------------------------------------------------------------------------------------


\begin{thebibliography}{}
\expandafter\ifx\csname natexlab\endcsname\relax\def\natexlab#1{#1}\fi
\providecommand{\url}[1]{\href{#1}{#1}}

\bibitem[{{Aschwanden} \& {Nightingale}(2005)}]{Aschwanden2005}
{Aschwanden}, M.~J., \& {Nightingale}, R.~W. 2005, \apj, 633, 499

\bibitem[{{Aschwanden} {et~al.}(2000){Aschwanden}, {Nightingale}, \&
  {Alexander}}]{Aschwanden2000}
{Aschwanden}, M.~J., {Nightingale}, R.~W., \& {Alexander}, D. 2000, \apj, 541,
  1059

\bibitem[{{Aschwanden} {et~al.}(2008){Aschwanden}, {Nitta}, {Wuelser}, \&
  {Lemen}}]{Aschwanden2008}
{Aschwanden}, M.~J., {Nitta}, N.~V., {Wuelser}, J.-P., \& {Lemen}, J.~R. 2008,
  \apj, 680, 1477

\bibitem[{{Brooks}(2019)}]{Brooks2019}
{Brooks}, D.~H. 2019, \apj, 873, 26

\bibitem[{{Brooks} {et~al.}(2012){Brooks}, {Warren}, \&
  {Ugarte-Urra}}]{Brooks2012}
{Brooks}, D.~H., {Warren}, H.~P., \& {Ugarte-Urra}, I. 2012, \apjl, 755, L33

\bibitem[{{Brooks} {et~al.}(2013){Brooks}, {Warren}, {Ugarte-Urra}, \&
  {Winebarger}}]{Brooks2013}
{Brooks}, D.~H., {Warren}, H.~P., {Ugarte-Urra}, I., \& {Winebarger}, A.~R.
  2013, \apjl, 772, L19

\bibitem[{{Brooks} \& {Yardley}(2021)}]{Brooks2021}
{Brooks}, D.~H., \& {Yardley}, S.~L. 2021, Science Advances, 7, eabf0068

\bibitem[{{Culhane} {et~al.}(2007){Culhane}, {Harra}, {James}, {Al-Janabi},
  {Bradley}, {Chaudry}, {Rees}, {Tandy}, {Thomas}, {Whillock}, {Winter},
  {Doschek}, {Korendyke}, {Brown}, {Myers}, {Mariska}, {Seely}, {Lang}, {Kent},
  {Shaughnessy}, {Young}, {Simnett}, {Castelli}, {Mahmoud}, {Mapson-Menard},
  {Probyn}, {Thomas}, {Davila}, {Dere}, {Windt}, {Shea}, {Hagood}, {Moye},
  {Hara}, {Watanabe}, {Matsuzaki}, {Kosugi}, {Hansteen}, \&
  {Wikstol}}]{Culhane2007}
{Culhane}, J.~L., {Harra}, L.~K., {James}, A.~M., {et~al.} 2007, Sol. Phys.,
  243, 19

\bibitem[{{De Rosa} {et~al.}(2009){De Rosa}, {Schrijver}, {Barnes}, {Leka},
  {Lites}, {Aschwanden}, {Amari}, {Canou}, {McTiernan}, {R{\'e}gnier},
  {Thalmann}, {Valori}, {Wheatland}, {Wiegelmann}, {Cheung}, {Conlon},
  {Fuhrmann}, {Inhester}, \& {Tadesse}}]{DeRosa2009}
{De Rosa}, M.~L., {Schrijver}, C.~J., {Barnes}, G., {et~al.} 2009, \apj, 696,
  1780

\bibitem[{{Del Zanna}(2013)}]{DelZanna2013}
{Del Zanna}, G. 2013, \aap, 555, A47

\bibitem[{{Del Zanna} {et~al.}(2021){Del Zanna}, {Dere}, {Young}, \&
  {Landi}}]{DelZanna2021}
{Del Zanna}, G., {Dere}, K.~P., {Young}, P.~R., \& {Landi}, E. 2021, \apj, 909,
  38

\bibitem[{{Del Zanna} \& {Mason}(2003)}]{DelZanna2003}
{Del Zanna}, G., \& {Mason}, H.~E. 2003, \aap, 406, 1089

\bibitem[{{Fleishman} {et~al.}(2020){Fleishman}, {Gary}, {Chen}, {Kuroda},
  {Yu}, \& {Nita}}]{Fleishman2020}
{Fleishman}, G.~D., {Gary}, D.~E., {Chen}, B., {et~al.} 2020, Science, 367, 278

\bibitem[{{Freeland} \& {Handy}(1998)}]{Freeland1998}
{Freeland}, S.~L., \& {Handy}, B.~N. 1998, \solphys, 182, 497

\bibitem[{{Gary}(2001)}]{Gary2001}
{Gary}, G.~A. 2001, \solphys, 203, 71

\bibitem[{{Gopalswamy} {et~al.}(2012){Gopalswamy}, {Nitta}, {Akiyama},
  {M{\"a}kel{\"a}}, \& {Yashiro}}]{Gopalswamy2012}
{Gopalswamy}, N., {Nitta}, N., {Akiyama}, S., {M{\"a}kel{\"a}}, P., \&
  {Yashiro}, S. 2012, \apj, 744, 72

\bibitem[{{Judge} {et~al.}(2016){Judge}, {Hutton}, {Li}, \&
  {Brage}}]{Judge2016}
{Judge}, P.~G., {Hutton}, R., {Li}, W., \& {Brage}, T. 2016, \apj, 833, 185

\bibitem[{{Kamio} {et~al.}(2010){Kamio}, {Hara}, {Watanabe}, {Fredvik}, \&
  {Hansteen}}]{Kamio2010}
{Kamio}, S., {Hara}, H., {Watanabe}, T., {Fredvik}, T., \& {Hansteen}, V.~H.
  2010, \solphys, 266, 209

\bibitem[{{Klimchuk} {et~al.}(1992){Klimchuk}, {Lemen}, {Feldman}, {Tsuneta},
  \& {Uchida}}]{Klimchuk1992}
{Klimchuk}, J.~A., {Lemen}, J.~R., {Feldman}, U., {Tsuneta}, S., \& {Uchida},
  Y. 1992, \pasj, 44, L181

\bibitem[{{Kosugi} {et~al.}(2007){Kosugi}, {Matsuzaki}, {Sakao}, {Shimizu},
  {Sone}, {Tachikawa}, {Hashimoto}, {Minesugi}, {Ohnishi}, {Yamada}, {Tsuneta},
  {Hara}, {Ichimoto}, {Suematsu}, {Shimojo}, {Watanabe}, {Shimada}, {Davis},
  {Hill}, {Owens}, {Title}, {Culhane}, {Harra}, {Doschek}, \&
  {Golub}}]{Kosugi2007}
{Kosugi}, T., {Matsuzaki}, K., {Sakao}, T., {et~al.} 2007, Sol. Phys., 243, 3

\bibitem[{{Kuridze} {et~al.}(2019){Kuridze}, {Mathioudakis}, {Morgan},
  {Oliver}, {Kleint}, {Zaqarashvili}, {Reid}, {Koza}, {L{\"o}fdahl},
  {Hillberg}, {Kukhianidze}, \& {Hanslmeier}}]{Kuridze2019}
{Kuridze}, D., {Mathioudakis}, M., {Morgan}, H., {et~al.} 2019, \apj, 874, 126

\bibitem[{{Landi} {et~al.}(2020){Landi}, {Hutton}, {Brage}, \&
  {Li}}]{Landi2020}
{Landi}, E., {Hutton}, R., {Brage}, T., \& {Li}, W. 2020, \apj, 904, 87

\bibitem[{{Landi} {et~al.}(2021){Landi}, {Li}, {Brage}, \&
  {Hutton}}]{Landi2021}
{Landi}, E., {Li}, W., {Brage}, T., \& {Hutton}, R. 2021, \apj, 913, 1

\bibitem[{{Lang} {et~al.}(2006){Lang}, {Kent}, {Paustian}, {Brown}, {Keyser},
  {Anderson}, {Case}, {Chaudry}, {James}, {Korendyke}, {Pike}, {Probyn},
  {Rippington}, {Seely}, {Tandy}, \& {Whillock}}]{Lang2006}
{Lang}, J., {Kent}, B.~J., {Paustian}, W., {et~al.} 2006, \ao, 45, 8689

\bibitem[{{Li} {et~al.}(2015){Li}, {Grumer}, {Yang}, {Brage}, {Yao}, {Chen},
  {Watanabe}, {J{\"o}nsson}, {Lundstedt}, {Hutton}, \& {Zou}}]{Li2015}
{Li}, W., {Grumer}, J., {Yang}, Y., {et~al.} 2015, \apj, 807, 69

\bibitem[{{Li} {et~al.}(2016){Li}, {Yang}, {Tu}, {Xiao}, {Grumer}, {Brage},
  {Watanabe}, {Hutton}, \& {Zou}}]{Li2016}
{Li}, W., {Yang}, Y., {Tu}, B., {et~al.} 2016, \apj, 826, 219

\bibitem[{{Lin} {et~al.}(2000){Lin}, {Penn}, \& {Tomczyk}}]{Lin2000}
{Lin}, H., {Penn}, M.~J., \& {Tomczyk}, S. 2000, \apjl, 541, L83

\bibitem[{{Malanushenko} {et~al.}(2009){Malanushenko}, {Longcope}, \&
  {McKenzie}}]{Malanushenko2009}
{Malanushenko}, A., {Longcope}, D.~W., \& {McKenzie}, D.~E. 2009, \apj, 707,
  1044

\bibitem[{{Malanushenko} {et~al.}(2012){Malanushenko}, {Schrijver}, {DeRosa},
  {Wheatland}, \& {Gilchrist}}]{Malanushenko2012}
{Malanushenko}, A., {Schrijver}, C.~J., {DeRosa}, M.~L., {Wheatland}, M.~S., \&
  {Gilchrist}, S.~A. 2012, \apj, 756, 153

\bibitem[{{Morgan} \& {Druckm{\"u}ller}(2014)}]{Morgan2014}
{Morgan}, H., \& {Druckm{\"u}ller}, M. 2014, \solphys, 289, 2945

\bibitem[{{Nakariakov} \& {Ofman}(2001)}]{Nakariakov2001}
{Nakariakov}, V.~M., \& {Ofman}, L. 2001, \aap, 372, L53

\bibitem[{{Reale}(2014)}]{Reale2014}
{Reale}, F. 2014, Living Reviews in Solar Physics, 11, 4

\bibitem[{{Rimmele} {et~al.}(2020){Rimmele}, {Warner}, {Keil}, {Goode},
  {Kn{\"o}lker}, {Kuhn}, {Rosner}, {McMullin}, {Casini}, {Lin}, {W{\"o}ger},
  {von der L{\"u}he}, {Tritschler}, {Davey}, {de Wijn}, {Elmore}, {Fehlmann},
  {Harrington}, {Jaeggli}, {Rast}, {Schad}, {Schmidt}, {Mathioudakis},
  {Mickey}, {Anan}, {Beck}, {Marshall}, {Jeffers}, {Oschmann}, {Beard},
  {Berst}, {Cowan}, {Craig}, {Cross}, {Cummings}, {Donnelly}, {de Vanssay},
  {Eigenbrot}, {Ferayorni}, {Foster}, {Galapon}, {Gedrites}, {Gonzales},
  {Goodrich}, {Gregory}, {Guzman}, {Guzzo}, {Hegwer}, {Hubbard}, {Hubbard},
  {Johansson}, {Johnson}, {Liang}, {Liang}, {McQuillen}, {Mayer}, {Newman},
  {Onodera}, {Phelps}, {Puentes}, {Richards}, {Rimmele}, {Sekulic}, {Shimko},
  {Simison}, {Smith}, {Starman}, {Sueoka}, {Summers}, {Szabo}, {Szabo},
  {Wampler}, {Williams}, \& {White}}]{Rimmele2020}
{Rimmele}, T.~R., {Warner}, M., {Keil}, S.~L., {et~al.} 2020, \solphys, 295,
  172

\bibitem[{{Si} {et~al.}(2020){Si}, {Brage}, {Li}, {Grumer}, {Li}, \&
  {Hutton}}]{Si2020}
{Si}, R., {Brage}, T., {Li}, W., {et~al.} 2020, \apjl, 898, L34

\bibitem[{{Ugarte-Urra} {et~al.}(2019){Ugarte-Urra}, {Crump}, {Warren}, \&
  {Wiegelmann}}]{Ugarte2019}
{Ugarte-Urra}, I., {Crump}, N.~A., {Warren}, H.~P., \& {Wiegelmann}, T. 2019,
  \apj, 877, 129

\bibitem[{{Van Doorsselaere} {et~al.}(2008){Van Doorsselaere}, {Nakariakov},
  {Young}, \& {Verwichte}}]{Vandoorsselaere2008}
{Van Doorsselaere}, T., {Nakariakov}, V.~M., {Young}, P.~R., \& {Verwichte}, E.
  2008, \aap, 487, L17

\bibitem[{{Warren} {et~al.}(2008){Warren}, {Ugarte-Urra}, {Doschek}, {Brooks},
  \& {Williams}}]{Warren2008}
{Warren}, H.~P., {Ugarte-Urra}, I., {Doschek}, G.~A., {Brooks}, D.~H., \&
  {Williams}, D.~R. 2008, \apjl, 686, L131

\bibitem[{{Warren} {et~al.}(2014){Warren}, {Ugarte-Urra}, \&
  {Landi}}]{Warren2014}
{Warren}, H.~P., {Ugarte-Urra}, I., \& {Landi}, E. 2014, \apjs, 213, 11

\bibitem[{{Winebarger} {et~al.}(2003){Winebarger}, {Warren}, \&
  {Mariska}}]{Winebarger2003}
{Winebarger}, A.~R., {Warren}, H.~P., \& {Mariska}, J.~T. 2003, \apj, 587, 439

\bibitem[{{Yang} {et~al.}(2020){Yang}, {Bethge}, {Tian}, {Tomczyk}, {Morton},
  {Del Zanna}, {McIntosh}, {Karak}, {Gibson}, {Samanta}, {He}, {Chen}, \&
  {Wang}}]{Yang2020}
{Yang}, Z., {Bethge}, C., {Tian}, H., {et~al.} 2020, Science, 369, 694

\end{thebibliography}
\end{document}